\documentclass[prb,reprint,twocolumn,preprintnumbers,amsmath,amssymb,showpacs,footinbib,superscriptaddress]{revtex4-1}

\usepackage[titletoc,toc,title]{appendix}
\usepackage{braket}
\usepackage[final]{feynmp}
\usepackage{ifpdf}
\usepackage{comment}
\usepackage{amsmath}
\usepackage{mathrsfs}
\usepackage{color}
\usepackage{bm}
\usepackage[font=small]{subcaption}
\usepackage[font=small]{caption}

\def\8{\infty}

\def\undertext#1{\vtop{\hbox{#1}\kern 1pt \hrule}}

\def\be{\begin{equation}}
\def\ee{\end{equation}}
\def\bea{\begin{eqnarray} & &}
\def\eea{\end{eqnarray}}

\def\rf#1{(\ref{#1})}

\def\rfs#1{Eq.~\rf{#1}}

\graphicspath{{./Figures/}}

\makeatletter
\def\endfmffile{%
	\fmfcmd{\p@rcent\space the end.^^J%
			end.^^J%
			endinput;}%
	\if@fmfio
		\immediate\closeout\@outfmf
	\fi
	\ifnum\pdfshellescape=\@ne
		\immediate\write18{mpost \thefmffile}%
	\fi}
\makeatother

\usepackage[demo]{graphicx}
\usepackage{dcolumn}
\usepackage{bm}
\usepackage{hyperref}

\newcommand{\beq}{\begin{equation}}
\newcommand{\eeq}{\end{equation}}

\begin{document}


\title{Dynamical quantum phase transitions in the random field Ising model}

\author{Victor Gurarie}
\affiliation{Department of Physics and Center for Theory of Quantum Matter, University of Colorado, Boulder, Colorado 80309, USA}

\begin{abstract}
Over the last few years it was pointed out that certain observables of time-evolving quantum systems may have singularities at certain moments in time, mimicking the singularities physical systems have when undergoing phase transitions. These were given the name of dynamical phase transitions. They were shown to exist in certain integrable (exactly solvable) quantum systems, and were seen numerically and experimentally in some models which were not integrable. The ``universality classes" of such singularities were not yet convincingly established, however. We  argue that random field Ising models feature singularities  in time 
which may potentially be present in a wider variety of quantum systems, in particular in those which are many body localized, and describe these singularities in detail analytically. 
\end{abstract}

\pacs{
}
\maketitle



Recently it has been proposed that some quantum systems evolving in time can have certain observables whose dependence on time is not analytic \cite{Heyl:2013fy}. One such observable is the ``return probability" related to the Loschmidt echo. Given an initial state $\left| \psi_0 \right>$ which is a ground state of a Hamiltonian $H_0$,  one could define the following observable 
\be Z(t) = \left< \psi_0 \right| e^{-i H t} \left| \psi_0 \right>,
\ee
where $H \not = H_0$. 
It is closely related to the trace of the evolution operator ${\rm tr} \, e^{-i H t}$, and to the partition function of the system ${\rm tr} \, e^{-H/T}$.  Indeed, expanding in the eigenstates of the Hamiltonian $\left| \psi_0 \right> = \sum_\alpha c_\alpha \left| \psi_\alpha \right>$, such that $H \left| \psi_\alpha \right> = E_\alpha \left| 
\psi_\alpha \right>$, we find
\be \label{eq:echo}  Z(t) = \sum_\alpha e^{-i E_\alpha t} \left| c_\alpha \right|^2 .
\ee
For some choice of $\left| \psi_0 \right>$ the coefficients $c_\alpha$ may happen to be all equal to each other. Even if not, the sum in \rfs{eq:echo} closely resembles that in the definition of thermal partition functions (in fact, \rfs{eq:echo} can be shown \cite{Silva2018} to be always equivalent to a partition function of some related system upon identifying $t=-i/T$). 
Partition functions are known to be non-analytic in $T$, signifying presence of phase transitions. It was proposed that $Z(t)$ may likewise be non-analytic in $t$. The physical significance of these singularities can be debated, but by now there is no doubt that they exist and can be measured\cite{Jurcevic:2017be,Flaschner:2018ch}. The ``universality classes" of these singularities have not yet been convincingly established. Most of the singularities discussed so far correspond to the discontinuities in the derivate $\partial \ln \left| Z \right|^2/  \partial t$,
resembling first order transitions in statistical physics. 

In this paper we demonstrate the existence of singularities in certain random models, of the new universality class $\partial \ln \left| Z \right|^2/  \partial t \sim \ln \left| t-t_0 \right|$. 
We argue that they should manifest themselves in a variety of random systems with Poisson level statistics, although probably not in the most generic many body localized systems.

We begin the discussion by reviewing the established facts, starting with the 1D classical Ising model\cite{Heyl:2015fm}. Its Hamiltonian is given by
\be \label{eq:1DI} H = -J \sum_{n=1}^N \sigma^z_n \sigma^z_{m+1},
\ee
with $\sigma^z_{N+1} \equiv \sigma^z_1$ (here and below, $\sigma^{x,y,z}$ denote Pauli matrices, while $\sigma=\pm 1$ will be the eigenvalues of $\sigma^z$). We could imagine the quench scenario where   the system is initially in the state $\left|\psi_0 \right>$ which is the ground state of a different Hamiltonian \be \label{eq:h0} H_0 = 
-\gamma \sum_{n=1}^N \sigma^x_n. \ee In other words, \be \label{eq:initgs} \left| \psi_0 \right> =\frac{1}{2^{N/2}} \prod_{n=1}^N
\sum_{\sigma= \pm 1} \left| \sigma \right>.\ee
 At a certain moment of time, the Hamiltonian suddenly changes (is quenched) to become $H$ of \rfs{eq:1DI}. Then
 $Z(t)$ can be calculated in a straightforward way
 \begin{eqnarray} \label{eq:1DIsing} Z(t) &=& \frac{1}{2^N} \sum_{\sigma=\pm 1} e^{i t J \sum_{n=1}^N \sigma_n \sigma_{n+1}} = \cr &&
 \left( \cos(Jt) \right)^N + \left(i \sin(J t ) \right)^N.
 \end{eqnarray}
In the large $N$ limit $F=\ln \left(\left| Z \right|^2 \right)/N$, the dynamical equivalent of free energy of the system, exhibits a nonanalytic behavior
\be F =\left\{  \begin{matrix}  \ln \left(\cos^2(tJ) \right), \ J t \in [-\pi/4, \pi/4] + \pi m, \cr
\ln \left( \sin^2(tJ) \right), \ J t \in [\pi/4,3\pi/4] + \pi m. \end{matrix} \right.
\ee
Here $m$ is an arbitrary integer. Specifically, $F$ has discontinuities in its derivative with respect to $t$ at points in time $J t_m = \pi/4+\pi m/2$, corresponding to a sort of the ``first order" singularities in $Z(t)$ of the 1D 
Ising model.

One should note that the thermal equivalent of \rfs{eq:1DI} gives
\begin{eqnarray} Z(T) &=& \sum_{\sigma=\pm 1} e^{\frac J T \sum_{n=1}^N \sigma_n \sigma_{n+1}} =
\cr &&
\left(2  \cosh \left(J/T\right) \right)^N + \left(2 \sinh(J/T) \right)^N.
\end{eqnarray}
This quantity is analytic in $t$ at large $N$, because $\left| \cosh(J/T) \right| > \left| \sinh(J/T) \right|$, signifying the absence of conventional  thermal phase transitions in a one dimensional Ising model (or in any other one dimensional system). 

Quite remarkably, $Z(t)$ defined for more general quantum systems also appear to have singularities at certain moments in time. For example, given a 1D transverse field Ising model\cite{Heyl:2013fy}
\be  \label{eq:1DT} H = - J \sum_{n+1}^N \sigma^z_n \sigma^z_{n+1} - \gamma \sum_{n=1}^N \sigma^x_n,
\ee
choosing $H_0$ to be $H$ with some choice of the parameters $J$, $h$, with $\left| \psi_0 \right>$ its ground state, and evolving this state with the Hamiltonian $H$ with some other values of these parameters, $Z(t)$ can be shows to have singularities at certain moments of time $t_m$ as long as the initial and final values of $J$ and $\gamma$ belong to two different phases of the 1D transverse field Ising model. The example of the 1D classical Ising model then becomes a particular case of this with the initial $J=0$ and the final $\gamma=0$. 

Furthermore, given that 1D transverse field Ising model is equivalent to free fermions by Jordan-Winger transformation, this construction was generalized to other free fermion systems\cite{Vajna2015} with their $Z(t)$ shown to have similar singularities at certain times $t_m$ under the right parameter quench. 
All such solvable systems however are examples of exactly solvable  (integrable) models. 

One can wonder if the singularities of $Z(t)$ in the examples above survive if the system considered is not integrable. Indeed, the moments in time $t_m$ where the singularities occur are related to level spacing $J^{-1}$ in the example worked out above. A generic quantum system with $N$ degrees of freedom (such as $N$ spin-1/2's) will have level spacing of the order of $e^{-N}$, leading to singularities, if any, occurring at enormous times
$t \sim e^{N}$, becoming unobservable for large systems $N \rightarrow \infty$. Nevertheless, nonintegrable
generalizations of \rf{eq:1DT} were studied numerically and found to still have the singularities in time\cite{Karrasch:2013ey,Karrasch:2017,Helimeh2017,Halimeh2017a}. Furthermore, these singularities were measured in an experiment\cite{Jurcevic:2017be} for 1D transverse field Ising model with nonlocal interactions in space, which is also not integrable.  

On the other hand, as an example of a very generic non-integrable (chaotic) quantum model we could consider a random matrix theory\cite{Mehta} (RMT). The quantity of interest to us is nothing but  the spectral form factor of RMT
\be \label{eq:RMT} \left| Z_{\rm RMT} (t) \right|^2 = \sum_{\alpha \beta} e^{it \left(E_\alpha-E_\beta \right)},
\ee
where $E_\alpha$ are energy levels of a random matrix. 
The derivatives of the RMT form factors over $t$ are known to have a discontinuity at a certain critical value\cite{Mehta} of $t$. One should note however that usually one studies the average form factor, while we are interested in the typical form factor 
which could be represented as the average of the logarithm of this quantity. The form factor is known not to be self-averaging\cite{Haake2015}, and its typical structure is not completely understood. Recent studies also looked at this and related quantities in quantum chaotic models such as the SYK model\cite{Cotler:2017ft}. 

Instead of a most generic quantum system with the Wigner-Dyson level statistics, let us consider a system with a Poissonian level statistics. Those models routinely appear in the context of many body localization\cite{Aleiner2006}. The simplest such model is 1D the random field classical Ising model, with the Hamiltonian
\be \label{eq:rfim} H = -J \sum_{n=1}^N \sigma^z_n \sigma^z_{n+1} - \sum_{n=1}^N h_n \sigma^z_n.
\ee
Here $h_n$ are random independent variables distributed uniformly on the interval $h \in [-J, J]$ (precise form of the probability distribution, as we argue below, turns out not to be important). The thermodynamics of this model was extensively studied in the past and was arguably found to be unremarkable\cite{Bruinsma:1983cx}. Here we are not interested in its thermodynamics, however. We again envision putting the system in the ground state \rfs{eq:initgs} of the Hamiltonian \rfs{eq:h0}, and then quenching it to \rfs{eq:rfim}.
The Loschmidt echo is again proportional to the imaginary temperature partition function of \rfs{eq:rfim}, to give
\be \label{eq:partrandom} Z =\frac{1}{2^N} \sum_{\sigma=\pm 1} e^{i t J \sum_{n=1}^N \sigma_n \sigma_{n+1} + i t \sum_{n=1}^N h_n \sigma_n}.
\ee
 $\left| Z \right|^2$, when averaged over random fields, is time independent at large enough time. Indeed,
 \begin{eqnarray} \left<  \left|Z \right|^2  \right> &=&\frac{1}{2^{2N}} \sum_{ \substack {\sigma=\pm 1 \\  \mu=\pm 1} } \int_{-J}^J \prod_{n=1}^N \left[ \frac{dh_n}{2J} \times \right. \cr && \left. e^{itJ \left( \sigma_{n}
 \sigma_{n+1} - \mu_n \mu_{n+1} \right) + it h_n \left(\sigma_n - \mu_n \right)} \right]. 
 \end{eqnarray}
When $tJ \gg 1$, the integrals over $h_n$ give Kronecker deltas $\delta_{\sigma_n \mu_n}$, resulting
in 
\be \label{eq:ave}  \left< \left| Z(t) \right|^2 \right> = 2^{-N}.
\ee
The Fourier transform of \rfs{eq:ave} with respect to $t$ is the delta function, reflecting the fact that the energy levels of \rfs{eq:rfim} are not correlated (compare with \rfs{eq:RMT}), indicative of the Poisson statistics of the levels. 

However, the typical values of $F=\ln\left( \left| Z \right|^2 \right)/N $ display a totally different and far more interesting behavior. Fig,~\rf{introfig} shows this quantity plotted numerically for $N=5000$ spins, for a single realization of random $h_n$. 
\begin{figure}[tb]
\centering
\includegraphics[width=0.46\textwidth]{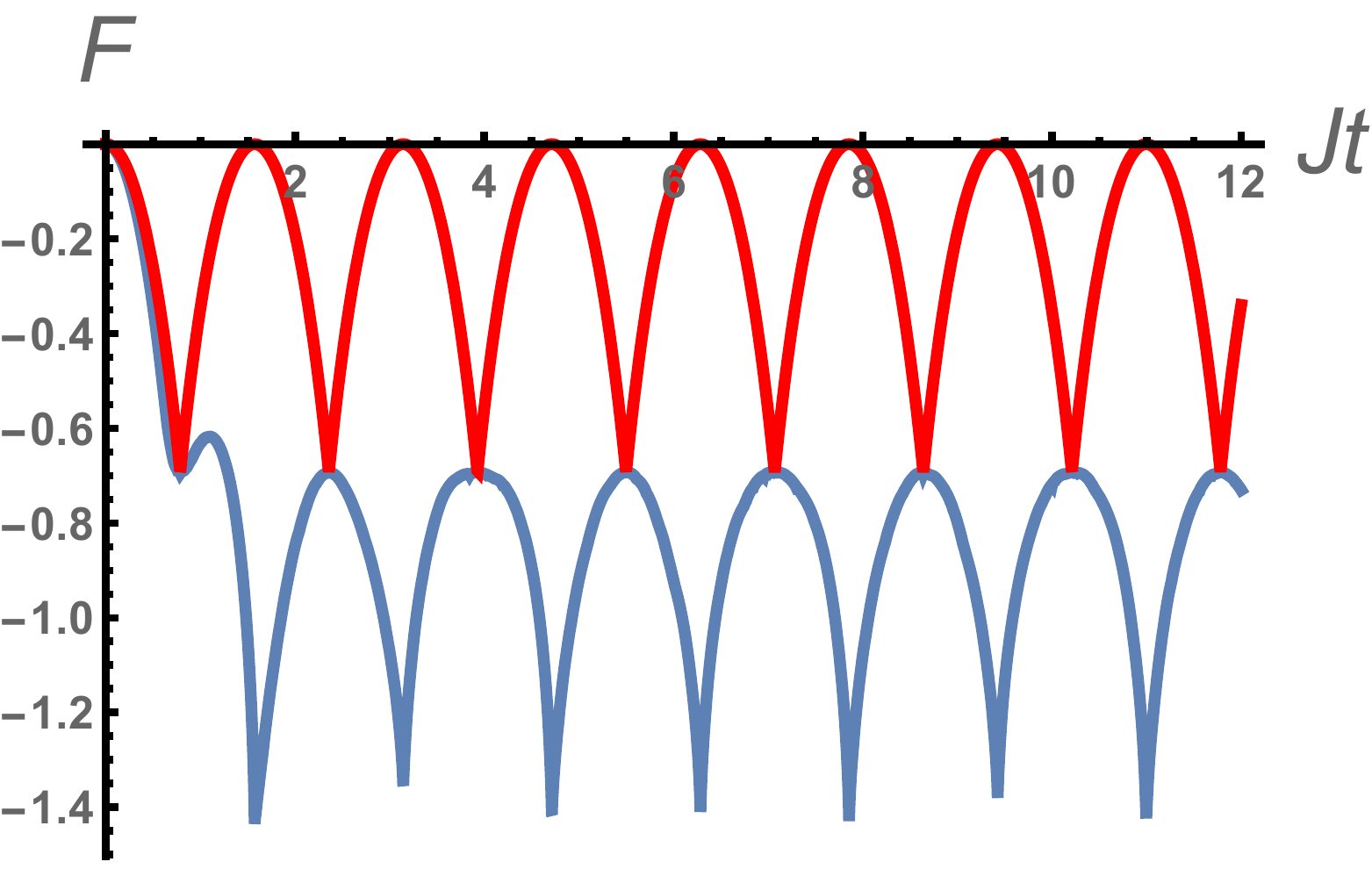}
\caption{The typical value of $F=\ln \left( \left|Z \right|^2 \right)/N$ from \rfs{eq:partrandom} computed for $N=5000$ spins and
random $h_n \in [-J, J]$ plotted as a function of $Jt$ (lower curve). The upper curve represents
the disorder free 1D Ising model \rfs{eq:1DIsing} shown for comparison. }
\label{introfig} 
\end{figure}
We see that unlike the average spectral form-factor this function displays criticalities which
superficially look similar to the disorder-free case, but as we now verify are qualitatively different from it.

To understand the nature of these singularities we observe (and justify later) that at large enough times  it is sufficient to think of $t h_n$ as being uniformly distributed on the interval $t h_n \in [-\pi, \pi]$. To confirm this, we plot the result of numerically evaluating
\be \label{eq:uniform}
Z =\frac{1}{2^N} \sum_{\sigma=\pm 1} e^{i t J \sum_{n=1}^N \sigma_n \sigma_{n+1} + i  \sum_{n=1}^N h_n \sigma_n}.
\ee
where $h_n$  are now randomly distributed on the interval $h_n \in [-\pi,\pi]$ in Fig.~\rf{introfig2}. The resulting curve coincides with \rfs{eq:partrandom} for large enough $t$.
\begin{figure}[b]
\centering
\includegraphics[width=0.46\textwidth]{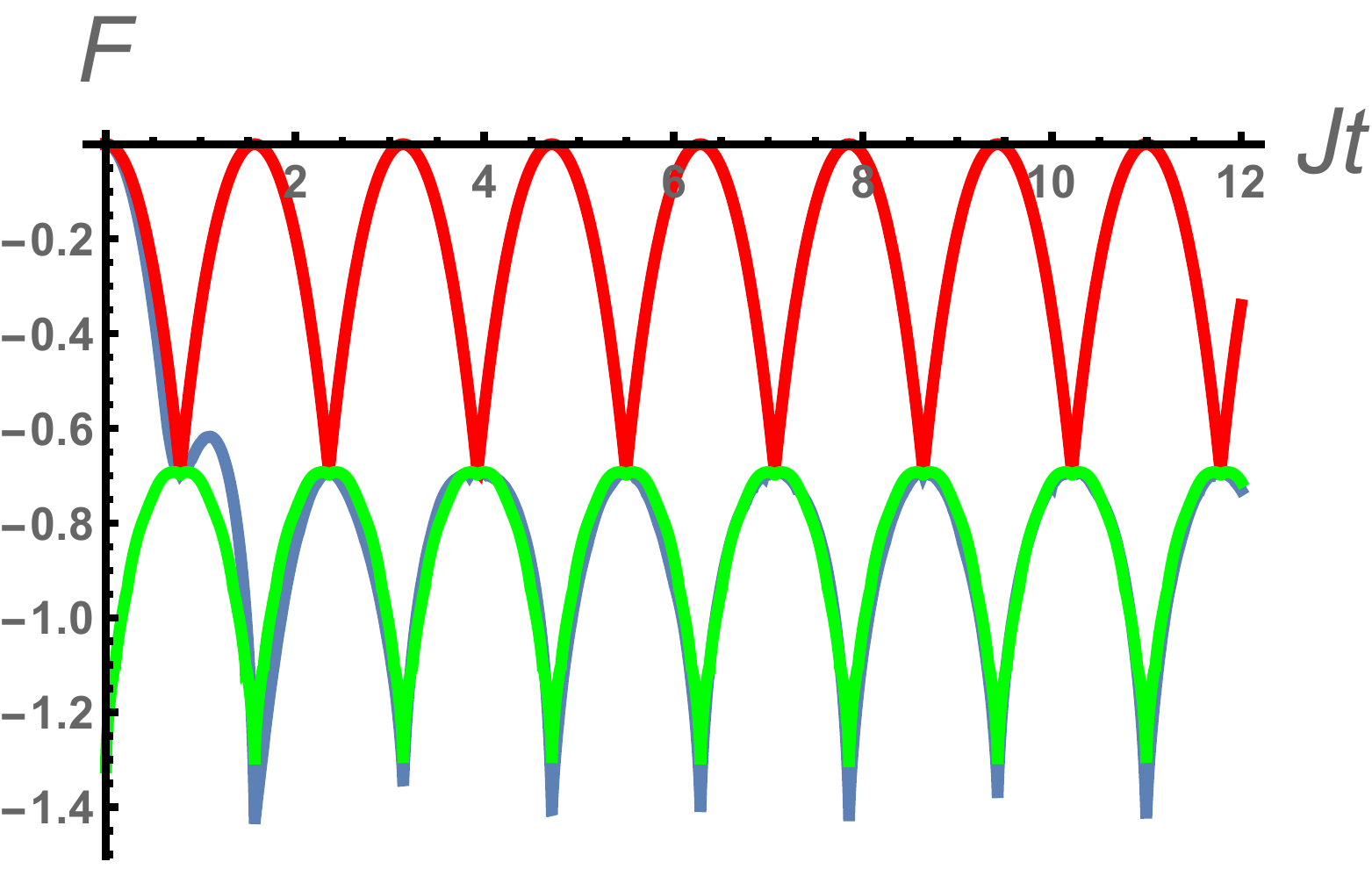}
\caption{Same as Fig.~\rf{introfig} but with the result of \rfs{eq:uniform} also shown. }
\label{introfig2} 
\end{figure}

From Fig.~(\ref{introfig} the singularities seem to occur at times $t_m = \pi m/(2 J)$. In fact, these are special times where the term proportional to $J$ in \rfs{eq:partrandom} can be set to zero without changing the value of $\left| Z \right|^2$. Choosing $t_m$ a multiple of $2 \pi /J$ for simplicity (the algebra is similar if slightly different for other values of $t_m$) we find
\be \label{eq:partcirc} Z = \frac{1}{2^N} \sum_{\sigma=\pm 1} e^{i \sum_{n=1}^N h_n \sigma_n} = \prod_{n=1}^N
\cos \left( h_n \right),
\ee
where from now on we take $h_n$ above to be uniformly distributed on the unit circle $h_n \in [-\pi, \pi]$
and no longer multiply them by $t$.
We expect $F$ to be a self averaging quantity. Averaging it over random $h_n$ gives
\be  F = \frac{1}{N} \int_{-\pi}^\pi \prod_{n=1}^N \left[  \frac{dh_n}{2 \pi} \right] \sum_{n=1}^n \ln \cos^2(h_n) = - \ln 4.
\ee 
This is consistent with the observed values of $F$ at $t_m=\pi m/(2 J)$ as seen in Fig.~\rf{introfig}.

Nearby these values of $t$, we substitute $t=t_m + J^{-1} \epsilon$ and expand $Z$ in powers of $\epsilon$
up to terms of the order $\epsilon^2$. This gives
\begin{eqnarray} Z &=& \frac{\cos^N(\epsilon) }{2^N}\sum_{\sigma=\pm 1} \prod_{n=1}^N \left( 1 + i  \tan (\epsilon) \sigma_{n} \sigma_{n+1} \right) e^{i h_n \sigma_n} \approx \cr && \prod_{n=1}^n \cos(h_n) \left( 1- i \epsilon \sum_{n=1}^N
\tan(h_n) \tan(h_{n+1})  - \right. \cr
&&  \epsilon^2 \sum_{n-m \ge 2} \tan(h_n) \tan(h_{n+1}) \tan(h_m) \tan (h_{m+1} ) + \cr
&& \left.  \epsilon^2 \sum_{n=1}^N \tan(h_n) \tan(h_{n+2}) \right). \label{eq:expansion}
\end{eqnarray}
This can be used to average $F$ over random $h_n$, accomplished by integrating it over $dh_n/(2\pi)$ over the interval $[-\pi, \pi]$ for each $h_n$. A convenient change of variables $\tan(h_n)=x_n$ brings the relevant expression to the form
\begin{eqnarray} && F \approx - \ln 4+ \frac{1}{\pi^N} \int_{-\infty}^\infty \prod_{n=1}^N \frac{dx_n}{1+x_n^2} \times
\cr && \ln \left( 1+ \epsilon^2 \sum_{n=1}^N \left(   x_n^2 x_{n+1}^2 + 2 x_n x_{n+2} + 2 x_n x_{n+1}^2 x_{n+2} \right) \right). \cr && \label{eq:integral}
\end{eqnarray}
So far everything appears to be analytic in $\epsilon$. However, the integral over $x_n$ makes the result nonanalytic. Indeed, expanding the logarithm in $\epsilon^2$ under the sign of integral we can easily see that the resulting integral is divergent, indicating that the result should be larger than $\epsilon^2$. Note that infinite $x$ corresponds to $h$ in the vicinity of $\pm \pi/2$, thus we predict that $J t \ge \pi/2$ for the singularities to appear in \rfs{eq:partrandom}.

We postpone the evaluation of \rfs{eq:integral} until the Appendix~\ref{sec:A}. Here we just state the result that
for small $\epsilon$ 
\be \label{eq:sing} F \approx -\ln 4 + \frac{4}{\pi} \epsilon \ln \frac 1 \epsilon.
\ee
This result is valid for $t=\pi m/(2J) + \epsilon$ for $\epsilon \ll 1/J$. 

To verify this we plot $F$ as a function of $\log \epsilon$ for small $\epsilon$, shown in Fig.~\rf{introfig3}. The result is consistent with \rfs{eq:sing}.

Thus despite appearing qualitatively similar in Fig.~\rf{introfig}, the singularities of the random field 1D Ising equation are much sharper than those for the nonrandom 1D Ising model, with the first derivative of $F$ diverging logarithmically as $t$ approaches either of the singularities.

\begin{figure}[tb]
\centering
\includegraphics[width=0.46\textwidth]{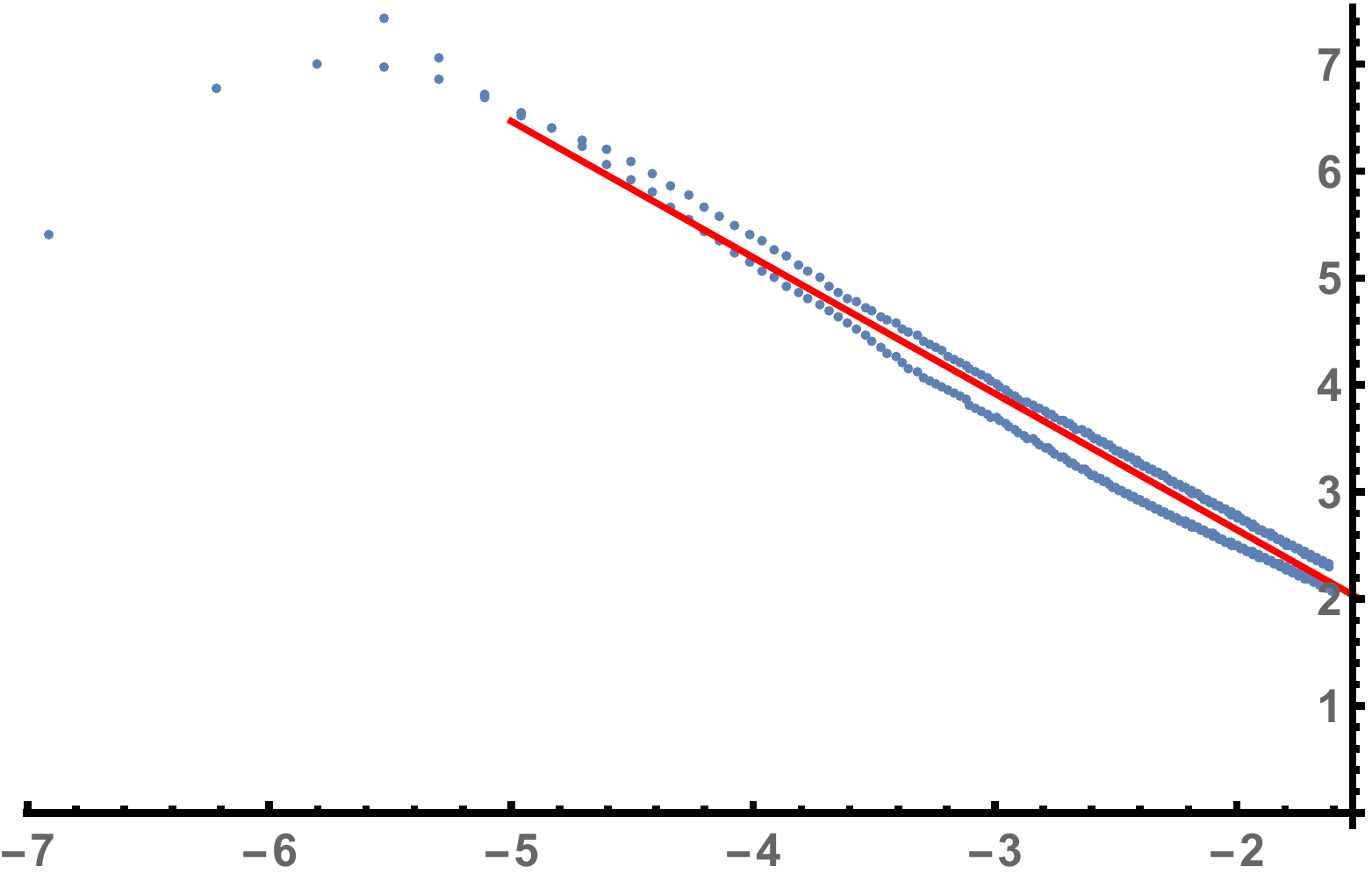}
\caption{Vicinity of singularity: $(F + \ln 4)/\epsilon$ is plotted as a function  of
$\ln \left| \epsilon \right|$ for \rfs{eq:uniform} at $N=5000$. The straight line has the slope of $-4/\pi$. Two sets of data correspond to the two signs of $\epsilon$.}
\label{introfig3} 
\end{figure}

A natural question is whether these singularities survive the addition of quantum terms to the Hamiltonian. For example, we could consider quenching \rfs{eq:h0} to the Hamiltonian
\be \label{eq:MBL} H = - J \sum_{n=1}^N \sigma_n^z \sigma_{n+1}^z - \sum_{n=1}^N h_n \sigma_n^z - \gamma
\sum_{n=1}^N \sigma_n^x,
\ee
with $h_n$ random as before. This model is not integrable and no good analytic methods exist to study 
its behavior. It is believed to have no quantum phase transitions at zero temperature\cite{Natterman1997} and to be many-body localized\cite{Huse2013}. As is well appreciated now, this implies that there exist a number of operators, called l-bits, in terms of which the Hamiltonian can be effectively diagonalized\cite{Imbrie:2016cs},
\be \label{eq:lbits} H = - \sum_{n=1}^N J^{(1)}_n \tau^z_n - \sum_{n=1}^N J^{(2)}_n \tau^z_n \tau^z_{n+1}+ \dots
\ee
where dots denote terms with a higher number of interacting spins, and $J^{(1)}_n$, $J^{(2)}_n$ all random. Such models however, where spin-spin interactions are now random, seem to wash out the singularities studied above, as is clear from Fig.~\rf{introfig4}. Thus a generic quantum random many-body localized model with energy levels obeying Poisson statistics would not have singularities of the type discussed here. 

\begin{figure}[tb]
\centering
\includegraphics[width=0.46\textwidth]{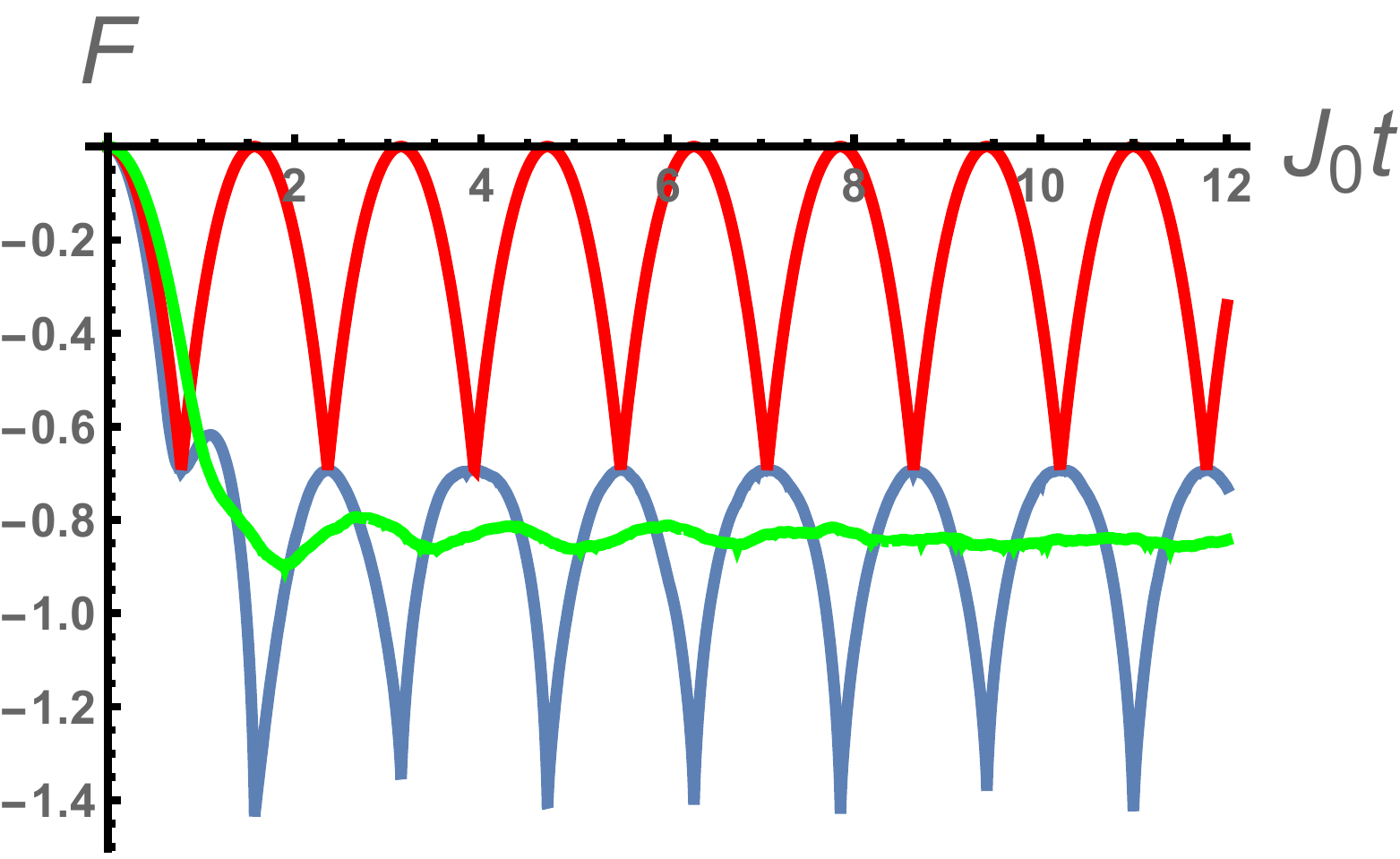}
\caption{Same as Fig.~\rf{introfig} but with additional line added representing $F$ for bonds $J$ randomly and uniformly distributed on the interval $[-J_0, J_0]$ in addition to $h_n$ uniform and random on the same interval.  }
\label{introfig4} 
\end{figure}

\rfs{eq:MBL} might not be as generic as \rfs{eq:lbits} with all coefficients random and uncorrelated, and indeed preliminary study of the exact diagonalization of the  Hamiltonian \rfs{eq:MBL} with $N=10$ was carried out and supported the hypothesis that for small enough $\gamma$ the singularities persist. Yet much larger numerical studies than what was already done need to be carried out to confirm this and be able to tell genuine singularities from crossovers numerically. We leave this for future work.

Nevertheless the behavior of the Loschmidt echo found here can be argued to be fairly generic. From \rfs{eq:expansion} leading to \rfs{eq:integral} it should be clear that a larger variety of classical models beyond 1D random field Ising model  should exhibit similar behavior. For example, consider classical 2D or 3D Ising models which are quenched from the initial paramagnetic state similar to \rfs{eq:initgs}, and whose Loschmidt echo is simply their partition function computed at imaginary temperature. When expanded in powers of $\epsilon$, random field averaging of $F$ leads to an integral broadly similar to \rfs{eq:integral} which should also produce the singularity $\epsilon \ln (1/\epsilon)$. 

Even more generally, one could observe that a large number of quantum systems can be thought of as consisting of fermionic ``quasiparticles" with the energy spectrum $\epsilon_\alpha$ and quasiparticle 
occupation numbers $n_\alpha = 0, 1$. The energy of such a system is
\be E = \sum_{\alpha} \epsilon_\alpha n_\alpha + J \sum_{\alpha \beta} n_\alpha n_\beta,
\ee where $J$ can be thought of as being quasiparticle type independent. Fermi liquids could be examples of such systems. Such systems have Poisson level statistics, as opposed to other ``more generic" quantum system whose levels obey Wigner-Dyson statistics.  It should be clear from the preceding discussion that all such models should have Loschmidt echo having the singularities of the type described here, as long as $J$ does not itself depend on $\alpha$ and $\beta$ in some random fashion. This construction gives a rather generic realization of the models considered here.

The author would like to acknowledge support from the NSF Grant No. PHY-1521080. The work described here was initiated at the KITP Santa Barbara during the program ``Intertwined Order and Fluctuations in Quantum Materials" and was also supported in part by the National Science Foundation under the NSF Grant No. PHY-1748958. The work was subsequently completed at the Erwin Schr\"odinger Institute in Vienna during the program ``Quantum Paths". The authors is grateful to Rahul Nandkishore, Leo Radzihovsky, Leon Balents, Stefan Kehrein and Jad Halimeh for discussions in the course of this work.

\appendix
\section{Calculation of the integral \rfs{eq:integral}}
\label{sec:A}

We would like to evaluate 
\be \label{eq:eee} I_N=\frac{1}{\pi^N} \int_{-\infty}^\infty \prod_{n=1}^N \frac{dx_n}{1+x_n^2} 
 \ln \left( 1+ \epsilon^2 X \right),
\ee
where \be X = \sum_{n=1}^N \left(   x_n^2 x_{n+1}^2 + 2 x_n x_{n+2} + 2 x_n x_{n+1}^2 x_{n+2} \right).
\ee
and show that for small $\epsilon$ it is approximately equal to 
\be - \frac{4N}{\pi} \epsilon \ln \epsilon.
\ee We will be able to show this for even $N$, although  this result very likely holds for any $N$. Therefore, from now on we take $N$ to be  even. 

In anticipation of the answer, we will calculate
\be \alpha =- \lim_{\epsilon \rightarrow 0} \left[  \epsilon \frac{\partial^2 I_N}{\partial \epsilon^2} \right].
\ee
and show that 
\be \alpha = \frac{4N}{\pi}.
\ee
Carrying out the differentiation we find
\be  \label{eq:alpha}
\alpha = \lim_{\epsilon \rightarrow 0} \frac{2}{\pi^N} \int_{-\infty}^\infty \prod_{n=1}^N \frac{dx_n}{1+x_n^2} 
\frac{ \epsilon X  \left( \epsilon^2 X - 1 \right)}{\left( 1+ \epsilon^2 X \right)^2}. \ee

At this point it is convenient to change odd labelled integration variables $x_{2n-1}$ according to 
\be \label{eq:change} x_{2n-1}=\frac {z_n}{\epsilon}. \ee We observe that 
\be \frac{1}{\pi}    \frac{dx_{2n-1}}{1+x_{2n-1}^2} = \frac 1 \pi 
\frac{\epsilon dz_n}{\epsilon^2+z_n^2}.
\ee
In turn, for small $\epsilon$, we can take advantage of the expansion
\be \label{eq:expd} \frac 1 \pi \frac{\epsilon }{\epsilon^2+z_n^2} \approx \delta(z_n) + \frac{\epsilon}{\pi z_n^2}.
\ee
This can for example be derived by Fourier transforming this expression with respect to $z_n$ obtaining $e^{-\epsilon \left|k \right|}$, where $k$ is the variable conjugate to $z_n$. Expanding in powers of $\epsilon$ and transforming back, we obtain \rfs{eq:expd}.
When applied to \rfs{eq:alpha} this becomes, with the convenient relabeling $x_{2n}=y_n$,
\begin{eqnarray} \alpha &=& 2 \lim_{\epsilon \rightarrow 0} \frac{1}{\epsilon \pi^{N/2}} \int_{-\infty}^\infty \prod_{n=1}^{N/2} dz_n \left[ \delta(z_n) + \frac{\epsilon}{\pi z_n^2} \right] \times \cr
&& \prod_{n=1}^{N/2} \frac{dy_{n}}{1+y_n^2} \frac{Y (Y-1)}{\left( 1+ Y \right)^2}. \label{eq:expande}
\end{eqnarray}
Here
\be Y = \sum_{n=1}^{N/2} \left( z_n^2 (y_n+y_{n+1})^2 + 2 z_n \left( y_n^2+1 \right) z_{n+1} +
2 \epsilon^2 y_n y_{n+1} \right).
\ee
The term in $Y$ proportional to $\epsilon^2$ is small and can be dropped. The rest of the integral can be calculated as an expansion over $\epsilon$. The term where delta functions are employed to do all 
integrals over $z_n$ can be seen to be zero. The first non-vanishing term is the one where the
integral over one of $z_n$ is done with the help of the second term in the square brackets  of 
\rfs{eq:expande}. There are $N/2$ such terms, one for each $z_n$, and they are all identical. 
They give
\begin{eqnarray} \alpha &=& \frac{N}{\pi^3} \int_{-\infty}^\infty \frac{dz dy_1 dy_2}{(1+y_1^2)(1+y_2^2)}
\times \cr &&
\frac{\left(y_1+y_2\right)^2 \left( z^2 \left(y_1+y_2 \right)^2-1\right)}{\left( 1+ z^2 \left(y_1+y_2 \right)^2\right)^2}.
\end{eqnarray}
To compute this integral, care needs to be taken because if $y_1+y_2=0$, then the integral over $z$ is divergent. Therefore, first one has to integrate over $y_1$ and $y_2$, and only then over $z$. The integrals over $y_1$ and $y_2$ give
\be \alpha = \frac{4N}{\pi} \int_{-\infty}^\infty \frac{1}{\left( 1+ 2 \left| z \right| \right)^2}.
\ee Doing this integral results in
\be \alpha = \frac{4N}{\pi},
\ee
which is the advertised result.

Finally, one may worry that terms higher order in $\epsilon$, dropped in the derivation of \rfs{eq:eee}, will not be small since upon \rfs{eq:change} $\epsilon$ may drop out of these higher order terms. However, all such terms are zero since they involve products of more than one $z_n$ and vanish thanks to the delta functions in 
\rfs{eq:expande}.


\bibliography{RandomField}


\end{document}